\documentclass[11pt]{article}
\usepackage{graphicx, epsfig}
\begin{document}

\title{Rapidity dependent $K/\pi$ ratios in Au+Au collisions at $\sqrt{s_{NN}}$ = 62.4 GeV}

\author{I.C.Arsene\footnote{Also at Institute for Space Sciences, Bucharest, Romania} for BRAHMS collaboration \\
  Department of Physics\footnote{e-mail: i.c.arsene@fys.uio.no}, University of Oslo, Norway
}

\maketitle

\begin{abstract}
We report on measurements of identified particle yields from Au-Au collisions 
at $\sqrt{s_{NN}} = 62.4$ GeV made with the BRAHMS spectrometer. Here we will concentrate on the
charged $K/\pi$ ratios as function of rapidity and baryo-chemical potential. We find that the
$K/\pi$ ratios measured at different rapidities in the analysed dataset have a common dependence
with the same ratios measured in mid-rapidity at SPS energies when plotted as function of
the $\bar{p}/p$ ratio. The theoretical models used for comparison, UrQMD and AMPT, give a reasonable
description of the particle yields at mid-rapidity but fail to do so for the $K/\pi$ ratios at forward rapidity.
\end{abstract}


The strange particle production in nucleus-nucleus and nucleon-nucleon collisions has 
been studied extensively. Experimentally, at AGS energies 
it has been observed that the strangeness ratio (\textit{i.e.} the number of strange 
particles divided by the number of pions) sharply increases with the beam energy in
A-A collisions \cite{agsKpiEnergy}. In the SPS experiments \cite{spsKpiEnergy} the strangeness ratio reached a maximum value 
at $E_{LAB} = 30A$GeV and continued with a very weak dependence on beam energy up to the highest 
RHIC energy. It has been also observed that the strangeness ratio increases with the
size of the colliding system \cite{agsKpiSystemSize}. At AGS energies, the strangeness enhancement in A-A collisions 
was explained phenomenologically by cascade models from rescatterings with heavy baryon resonances. 
The saturation of the strangeness ratio seen
at the SPS experiments and later at RHIC was interpreted by some\cite{qgpSignal} as a signal of the transition
to a deconfined state of nuclear matter.

The $K^{+}/\pi^{+}$ and $K^{-}/\pi^{-}$ ratios have a slightly 
different excitation function than the strangeness ratio. At AGS and lower SPS energies,
the $K^{+}/\pi^{+}$ ratio at mid-rapidity is approximately proportional to the strangeness ratio but at
$E_{LAB} = 30A$GeV exihibits a "horn"-like shape and decreases slowly with increasing
collision energy. The overall behaviour has been accounted for by effects like isospin, increasing yields of
(anti)strange baryons which get an increasing with energy share of the strange quarks created in the
collision \cite{kpiIsospin}, transition from baryon to meson dominated matter and possibly a
phase transition.

\begin{figure}[hpt]
\begin{center}
\begin{tabular}{c @{} c}
\includegraphics[height=2.0in]{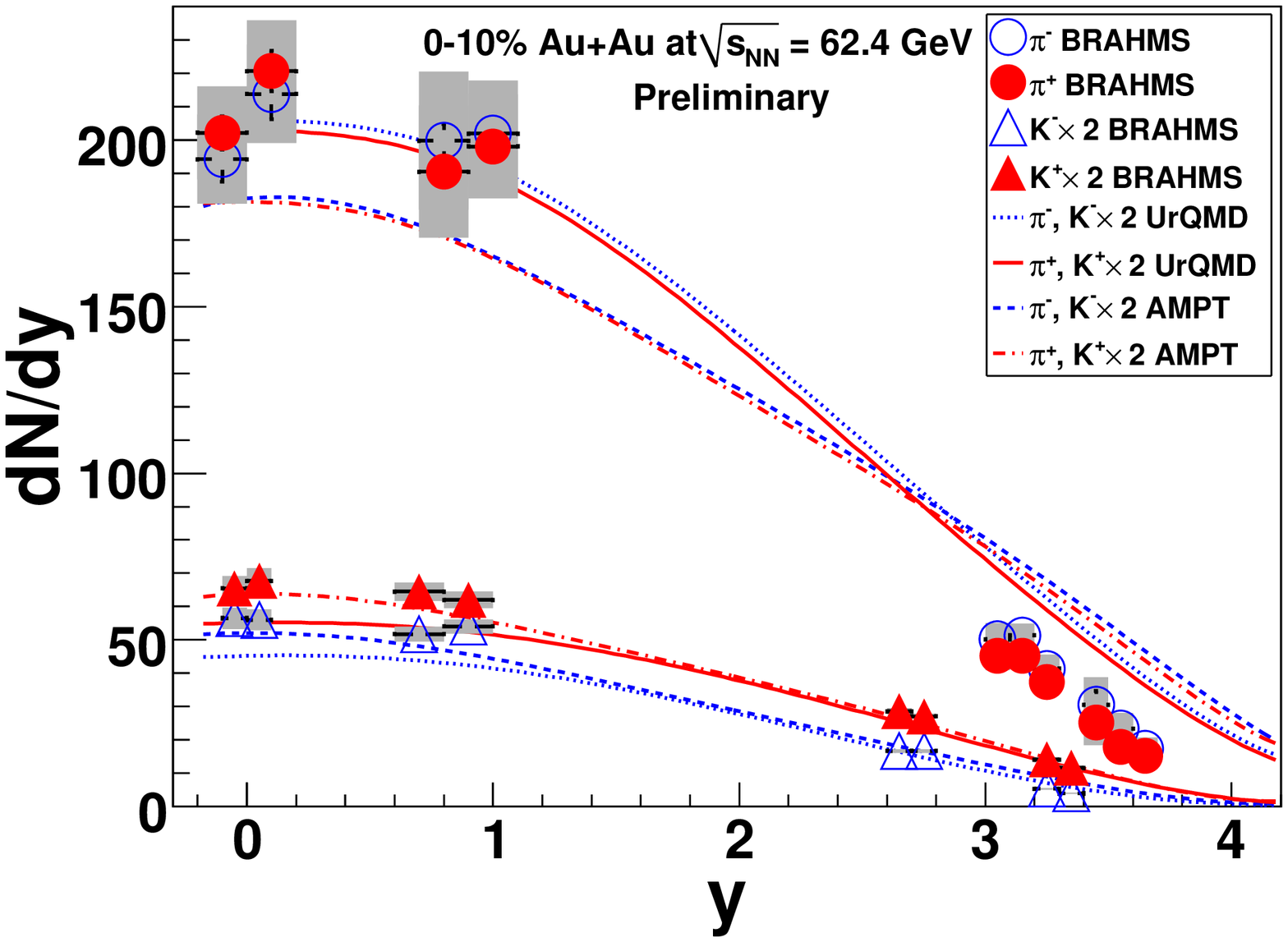} &
\includegraphics[height=2.0in]{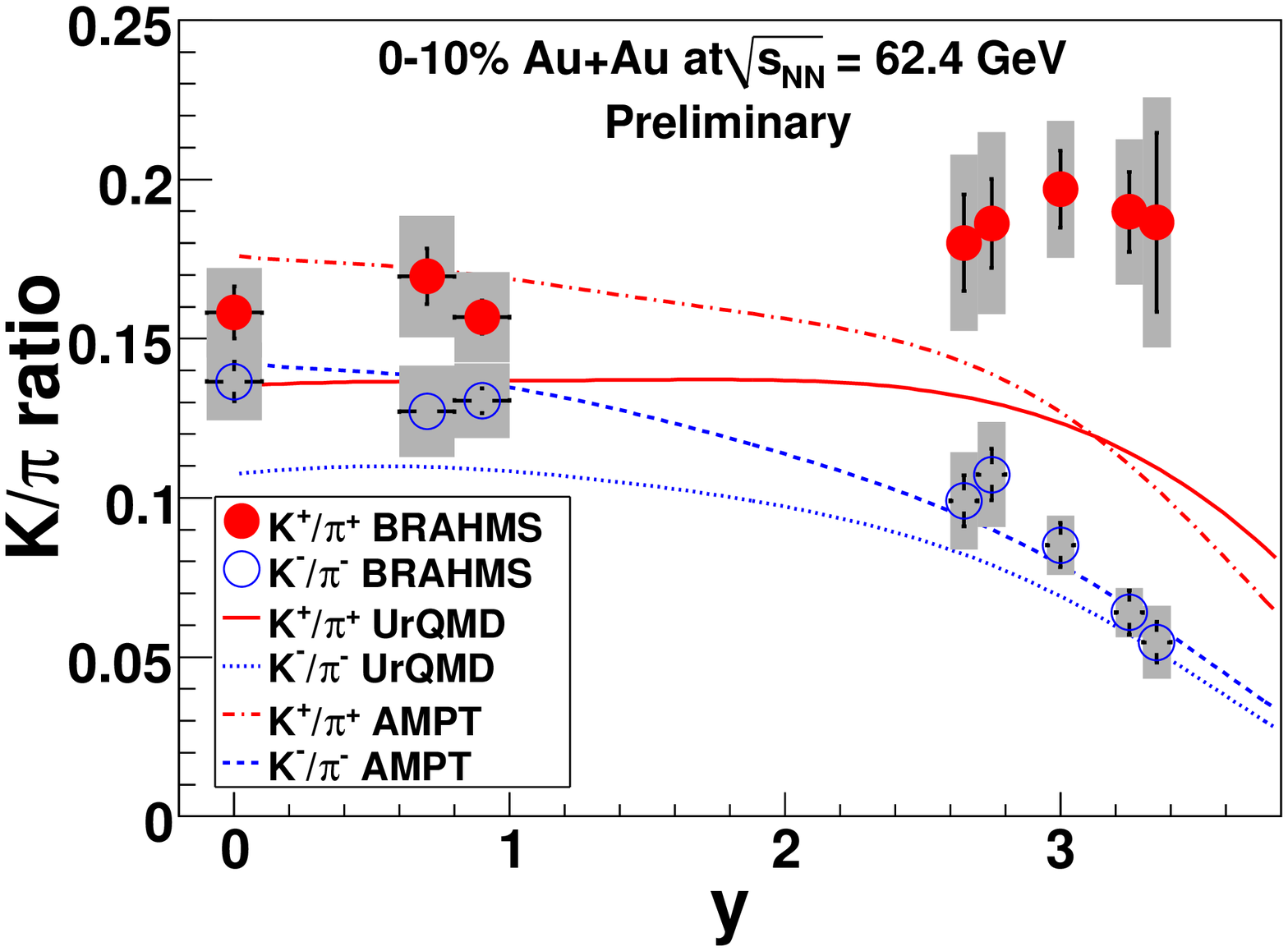} \\
(a) & (b)
\end{tabular}
\caption{Left: Rapidity density for charged pions and kaons in 0-10\% Au+Au collisions as a function of rapidity. 
Right: $K/\pi$ ratios as a function of rapidity. 
The data points in both figures are BRAHMS measurements. The error bars are statistical and the shaded rectangles
are systematic errors from the yield extrapolation. The curves are UrQMD \cite{urqmd} and AMPT \cite{ampt} model calculations. 
The solid curves denote the $K^{+}$, $\pi^{+}$ yields and $K^{+}/\pi^{+}$ ratio while the dotted curve
stands for the $K^{-}$, $\pi^{-}$ yields and $K^{-}/\pi^{-}$ ratio obtained with UrQMD.
The dot-dashed line represents the $K^{+}$, $\pi^{+}$ yields and $K^{+}/\pi^{+}$ ratio and
the dashed line represents the $K^{-}$, $\pi^{-}$ yields and $K^{-}/\pi^{-}$ ratio obtained with AMPT.}
\label{dndy}
\end{center}
\end{figure}

The BRAHMS \cite{brahmsNIM, brahmsWhite} experiment consists of two rotatable spectrometer arms for 
particle identification and momenta measurement. It contains also a set of global detectors 
for bulk event characterization. 
BRAHMS has a very wide phase space coverage with very good particle
identification. In the $\sqrt{s_{NN}} = 62.4$ GeV dataset BRAHMS identified charged 
particles up to rapidity 3.4 which is less than one rapidity unit away from the beam. The
$\pi/K$ separation can be made up to $p\approx2.5$GeV/$c$ near mid-rapidity and $p\approx20$GeV/$c$ at $y>2$. The protons
are well identified up to $p\approx3.0$GeV/$c$ near mid-rapidity and $p\approx30$GeV/$c$ at $y>2$.

The spectra were corrected for tracking and PID efficiencies, acceptance, in-flight
weak decays, multiple scattering and hadronic interactions. 
The invariant $p_{T}$ spectra was
extracted in different rapidity windows and fitted with several theoretically motivated functions
in order to extract the integrated yields. The charged pion spectra was fitted best 
with a power law function $A(1+p_T/p_0)^{-B}$
at mid-rapidity and with an $m_T$ exponential function $Aexp(-m_T/T)$ at forward rapidity $y\geq3.0$.
The kaon and proton spectra were fitted equally well with the $m_T$ exponential function and with
an $m_T$ Boltzmann distribution over the entire rapidity range \cite{brahmsSpectra}.
The resulting $p_T$ integrated yields are shown in Fig.\ref{dndy}(a) as a function of rapidity
and includes the systematic errors that in part comes from the extrapolations to low $p_T$.
The data are compared to the predictions of two microscopic
models, UrQMD v2.2 \cite{urqmd} and AMPT v1.11 \cite{ampt}. Both of the models include a partonic initial stage followed by a Lund-type string 
fragmentation and hadronic rescatterings. At mid-rapidity there is some disagreement
between models but at forward rapidity it looks like the models converge. Both of the models 
overestimate the measured pionic yields at forward rapidity while they describe the kaon yield within errors.

In Fig.\ref{dndy}(b) we show the $K/\pi$ ratios dependence on rapidity. The ratios
are obtained by dividing the kaon and pion $p_T$ integrated yields. Because the
phase space covered by the two species is quite different at forward rapidity, for
the points at $y\approx2.7$ and $y=3.0$ we used linear interpolation to calculate the
pion and kaon yields respectively. We observe that the $K^{-}/\pi^{-}$
ratio drops from a value of $\approx 0.13$ at mid-rapidity to $\approx 0.05$
at $y=3.3$ while the corresponding positive ratio exhibits a slow increase from
$K^{+}/\pi^{+} \approx 0.16$ at $y = 0$ to a value of $\approx 0.2$ at $y = 3$.
Both UrQMD and AMPT models describe the rapidity behaviour of the $K^{-}/\pi^{-}$
ratio but fail to do so for the $K^{+}/\pi^{+}$ ratio.

\begin{figure}[hpt]
\begin{center}
\begin{tabular}{c @{} c}
\includegraphics[height=2.3in]{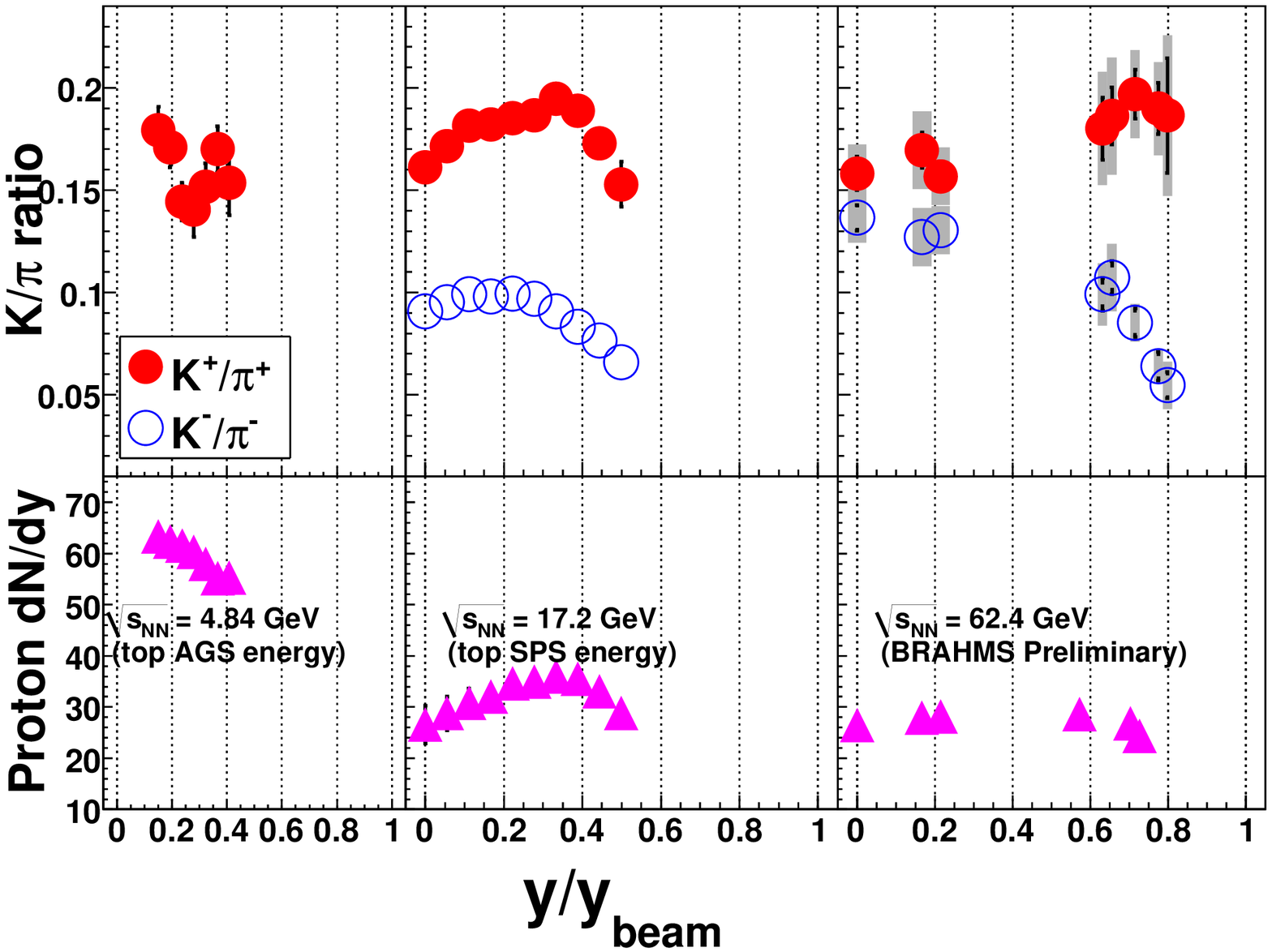} &
\includegraphics[height=2.3in, width=2.8in]{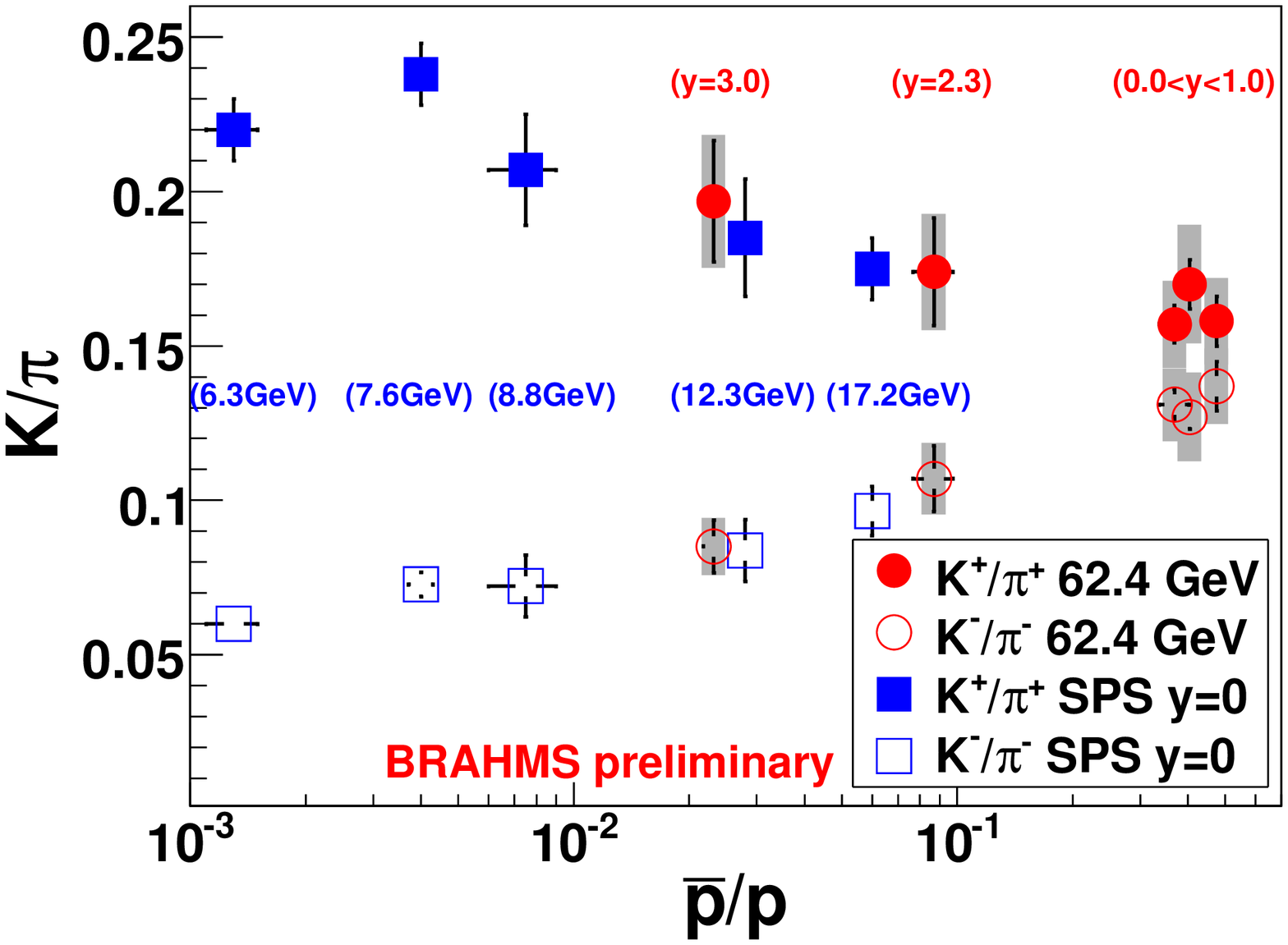} \\
(a) & (b)
\end{tabular}
\caption{Left: $K/\pi$ ratios and proton densities from central A+A collisions as a 
function of the normalized rapidity. The solid circles are for the $K^{+}/\pi^{+}$ ratio 
and the open circles for the $K^{-}/\pi^{-}$ ratio.
Right: $K/\pi$ ratios dependence on the $\bar{p}/p$ ratio in central A+A collisions from SPS [2] and
RHIC measurements. The solid and open symbols are the $K^{+}/\pi^{+}$ and $K^{-}/\pi^{-}$ ratios, respectively.
The squares are for the ratios measured at mid-rapidity in SPS experiments at (from left to right) 
$\sqrt{s_{NN}}=$6.3, 7.6, 8.8, 12.3 and 17.2 GeV. The circles denote the points measured by BRAHMS at 
$\sqrt{s_{NN}}=62.4$ GeV at different rapidities.}
\label{kpiComparison}
\end{center}
\end{figure}

In Fig.\ref{kpiComparison}(a) and (b) we show a comparison of our $K/\pi$
ratios measurements with the lower energy data. The 
first row of Fig.\ref{kpiComparison}(a) shows the rapidity dependence of the $K/\pi$ ratios
(left to right) at top AGS energy \cite{agsKpiSystemSize, agsProtons}, top SPS energy 
\cite{spsProtons} and BRAHMS at 62.4 GeV \cite{brahmsProtons}.
On the second row we show the corresponding proton density
distributions. In all the three situations we observe that the $K^{+}/\pi^{+}$ ratios
are enhanced compared to the ones measured in p+p or p+A experiments \cite{ppAgs, ppIsr}. The enhanced production of
strange kaons at AGS is believed to be the effect of additional scatterings of bound 
hadrons during the collision, \textit{e.g.} $\pi + N \rightarrow K^{+} + \Lambda$. The corresponding
anti-particle channel which creates $K^{-}$ and $\bar{\Lambda}$ gives a much smaller contribution
because of the very small anti-nucleon densities.
From SPS towards RHIC energies we observe a transition from baryon to meson dominated matter together
with the appearance of collision transparency. However, there is no sizeable change in the values of the
$K^{+}/\pi^{+}$ ratios at mid-rapidity. The large strangeness ratios measured at SPS energies were
interpreted as one of the signals to the formation of quark-gluon plasma.
From the SPS data at $E_{LAB} = 158$ GeV/c (see middle row in Fig.\ref{kpiComparison}(a)) 
we observe a correlation in the behaviour of the $K^{+}/\pi^{+}$ ratio with the proton rapidity density.
At $y\approx1.2$, where the proton density has a maximum, the $K^{+}/\pi^{+}$ peaks as well and this
might be due partly to the same hadronic channels which created extra strangeness at lower AGS energies.
At $\sqrt{s_{NN}} = 62.4$ GeV kaons and pions are measured up to $y\approx3.4$ which is very 
close to the beam rapidity ($y_{beam} \approx 4.2$).
The $K^{+}/\pi^{+}$ ratio is 0.16 at mid-rapidity and grows to a value of approximately 0.2 at rapidity 
$y\approx3$ where the fragmentation peak is located.


In Fig.\ref{kpiComparison}(b) we show the dependence of the $K/\pi$ ratios with the $\bar{p}/p$ ratio.
There is a common dependence of the $K/\pi$ ratios with the $\bar{p}/p$ ratio, whether measured
for different energies at mid-rapidity at SPS, or at different rapidities at 62.4 GeV. The $K^{-}/K^{+}$
ratio exhibits a similar feature.
Also the pionic and protonic yields in the range where the baryo-chemical potential from RHIC and SPS
overlap are approximately the same.
This suggests that the local system formed at high rapidity at RHIC (62.4 GeV) is chemically equivalent
with the system formed at SPS at mid-rapidity.

In this work we presented experimental results from the most central 10\% Au+Au collisions at
$\sqrt{s_{NN}} = 62.4$ GeV measured in the BRAHMS experiment. The rapidity dependence of the $K^{+}/\pi^{+}$
ratio shows higher values at forward rapidity than at mid-rapidity which is in contradiction
with the expectations of the UrQMD and AMPT microscopic transport models. We showed that there is a
common dependence of the $K/\pi$ and $K^{-}/K^{+}$ ratios with the $\bar{p}/p$ ratio when
looking at SPS results in mid-rapidity together with our results from different rapidity slices.
The nuclear mediums formed in nuclear collisions at mid-rapidity at SPS experiments and at forward rapidity
in RHIC(62.4 GeV) develop the same chemistry which seems to be driven by the baryo-chemical potential.

\end{document}